\documentclass[useAMS,usenatbib]{mn2e}
\usepackage{amsmath}
\usepackage{times}
\usepackage{graphicx}
\usepackage{aas_symbols}
\usepackage{graphicx}
\usepackage{epstopdf}
\usepackage{ulem}

\newcommand{\beq}{\begin{equation}}
\newcommand{\eeq}{\end{equation}}
\newcommand{\bea}{\begin{eqnarray}}
\newcommand{\eea}{\end{eqnarray}}
\newcommand{\gae}{\lower 2pt \hbox{$\, \buildrel {\scriptstyle >}\over {\scriptstyle
\sim}\,$}} 
\newcommand{\lae}{\lower 2pt \hbox{$\, \buildrel {\scriptstyle <}\over {\scriptstyle
\sim}\,$}}

\begin{document}

\title[Anisotropic minijets]{An anisotropic minijets model for the GRB prompt emission}

\author[Barniol Duran, Leng \& Giannios]{R. Barniol Duran$^{1}$\thanks{Email: rbarniol@purdue.edu, mleng@purdue.edu, dgiannio@purdue.edu}, M. Leng$^{1}$\footnotemark[1], D. Giannios$^{1}$\footnotemark[1] \\
$^{1}$Department of Physics and Astronomy, Purdue University, 525 Northwestern Avenue, West Lafayette, IN 47907, USA
}

\date{Accepted 2015 September 23. Received 2015 September 22; in original form 2015 September 8}

\pubyear{2015}

\maketitle

\begin{abstract}
In order to explain rapid light curve variability without invoking a variable source, several authors have proposed ``minijets" that move relativistically relative to the main flow of the jet.  Here we consider the possibility that these minijets, instead of being isotropically distributed in the comoving frame of the jet, form primarily perpendicular to the direction of the flow, as the jet dissipates its energy at a large emission radius. This yields two robust features. First, the emission is significantly delayed compared with the isotropic case. This delay allows for the peak of the afterglow emission to appear while the source is still active, in contrast to the simplest isotropic model.  Secondly, the flux decline after the source turns off is steeper than the isotropic case. We find that these two features are realized in gamma-ray bursts (GRBs): 1. The peak of most GeV light curves (ascribed to the external shock) appears during the prompt emission phase.  2. Many X-ray light curves exhibit a period of steep decay, which is faster than that predicted by the standard isotropic case.  The gamma-ray generation mechanism in GRBs, and possibly in other relativistic flows, may therefore be anisotropic.
\end{abstract}


\section{Introduction}
The light curves of Gamma-ray bursts (GRBs) and blazar jets in active galactic nuclei (AGNs) show variability on short time-scales.  Various models have been put forward to explain this feature.  Among them, the minijets model, also called ``jets-in-a-jet", ``fundamental emitters" or ``relativistic turbulence" model (e.g., \citealp{blandford02, lyutikovandblandford03, lyutikov06a, lyutikov06b, gianniosetal09, kumarandnarayan09, lazaretal09, narayanandkumar09, gianniosetal10, nalewajkoetal11, narayanandpiran12, giannios13}), provides a promising explanation. In this model, compact active regions of the relativistic jet experience relativistic motions {\it relative} to the comoving frame of the jet.  We will refer to these regions as minijets.  The minijets radiate and yield short time pulses, which explain the fast variability observed in these sources.

Ideally, one would like to constrain the various properties of the minijets, for example, their number, their velocity and their direction relative to the main flow of the jet with available observations (e.g., \citealp{gianniosetal09, narayanandkumar09}). In fact, recent AGN observations point out that the $\gamma$-ray bright blazars are preferentially viewed in a direction perpendicular to that of the main flow of the jet in the comoving frame of the jet \citep{savolainenetal10}. This is at odds with the expectation that the jet emission is isotropic in the rest frame of the jet, but instead points: (i) to the presence of minijets, and (ii) to the fact that most of the minijets are preferentially perpendicular to the jet propagation. In general, the direction of the minijets is anisotropic. The main objective of this Letter is to explore this possibility. 

Anisotropic emission has already been studied in the context of the GRB afterglow emission \citep{beloborodovetal11}, and two main properties were found.  First, anisotropic emission is delayed with respect of the isotropic one.  And second, after the source turns off, the flux declines much faster for an anisotropic source compared with an isotropic one \citep{narayanandkumar09}.  We find these two same features in the anisotropic minijet model and explore their consequences in the context of GRB observations.  

The main equations of the minijets model are presented in Section \ref{Section_2}.  In Section \ref{Section_3} we highlight the main two features that appear in the light curves if the direction of the minijets is anisotropic.  In Section \ref{Results} we present typical light curves for different levels of anisotropy. We connect our results within the context of GRBs in Section \ref{Application}, and finish with a short discussion and our conclusions in Section \ref{Discussion}.

\section{Minijets model} \label{Section_2}

GRBs are characterized by $t_{var}\ll t_{GRB}$, where $t_{var}$ is the average duration of a single spike in the prompt emission light curve of total duration $t_{GRB}$. One popular interpretation is that $t_{var}$ directly reflects changes in the central engine and corresponds to an emission distance $R_{diss}\sim \Gamma^2 ct_{var}$ (e.g. internal shocks model).  In an alternative view, the engine is active for $t_{GRB}$ ejecting a shell of thickness $\Delta=c t_{GRB}$. When the shell reaches a large distance $R_{diss}\sim \Gamma^2 c t_{GRB}$, it turns dissipative producing the GRB. The model can account for the short variability with ``minijets". Each of them beams the emission in the rest frame of the jet powering individual pulses while the overall duration of the event is $t_{GRB}\sim R_{diss}/\Gamma^2 c\sim \Delta/c$. Note that the two models are not mutually exclusive: minijets may take place even if $R_{diss}\sim \Gamma^2 ct_{var}$, while residual shocks take place out to distance $\sim \Gamma^2 c t_{GRB}$ and their emission may be anisotropic. 

Here we focus on a model with a large emission radius and explore different orientations for the minijets. We develop a Monte Carlo scheme to model minijets in a homogeneous jet of radius $R$, thickness $\Delta$ and opening angle $\theta_j$, moving towards the observer with velocity $v_j$ and corresponding Lorentz factor $\Gamma_j$. For simplicity, we assume all minijets are identical, and they all emit isotropically with the same luminosity $L_{mj}$ for a duration $\delta t_{mj}$ (in their rest frame, `mj' stands for the ``minijet" frame).

The velocity of the minijet is $v_{co}$ (Lorentz factor $\Gamma_{co}$), measured in the comoving frame of the jet. We randomly select the position of the minijet within the jet: its radius $r$ ($R-\Delta < r < R$) and angle $\Theta$ ($0<\Theta<\theta_j$), which is the angle subtended by the radial direction at the location of the minijet and the symmetry axis (given the symmetry of the problem, the azimuthal angle is not needed). For each minijet, we select two angles, measured in the comoving frame of the shell, which describe the direction of the minijet: $\theta'$ and $\phi'$, which are the polar and azimuthal angles with respect to the radial direction.  The angle $\theta'$ is selected from a distribution $P(\theta')$ such as $\int P(\theta') \,d\Omega' = 4\pi$.  We consider a distribution of minijets that is symmetric around $\pi/2$, i.e., $P(\theta') \propto (\sin \theta')^n$, with $n \geq 0$. For $n=0$ we recover an isotropic distribution with $P(\theta') =1$, and as $n$ increases the distribution peaks strongly around $\pi/2$, and the chance of minijets having $\theta' \sim \pi/2$ increases. 

In the lab frame, each minijet moves with Lorentz factor
\beq \label{LF_minijets}
\Gamma_{em} = \Gamma_j \Gamma_{co} (1 + v_j v_{co} \cos \theta'),
\eeq
and corresponding velocity $v_{em}$, at an angle $\theta$, given by
\beq \label{lab_frame_angle}
\tan \theta = \frac{v_{co} \sin \theta'}{\Gamma_j (v_{co} \cos \theta' + v_j)}.
\eeq
For the azimuthal direction $\phi = \phi'$. We set the arrival time of photons from a minijet located at $r=R$ along $\Theta = 0$ to be $t_0$.  Photons from other minijets will arrive later, at a time
\beq \label{time_delay}
t - t_0= \frac{R - r \cos \Theta}{c},
\eeq
where $c$ is the speed of light.

The angle between the direction of the minijet and a distant observer, $\alpha$, is $\cos \alpha = \sin \Theta \sin \theta \sin \phi + \cos \Theta \cos \theta$, and the Doppler factor is 
\beq \label{Doppler_factor_minijets}
\mathcal{D} = \frac{1}{\Gamma_{em} (1 - v_{em} \cos \alpha)}.
\eeq
Given the luminosity of the minijet $L_{mj}$ in its own rest frame, then the observed luminosity is
\beq
L_{obs} = L_{mj} \mathcal{D}^4.
\eeq
The observed duration of a flare powered by a minijet is
\beq
\delta t = \frac{\delta t_{mj}}{\mathcal{D}} \approx \frac{R}{\Gamma_j \Gamma_{co} c \mathcal{D}},
\eeq
where we approximated the duration of the minijet emission in its own frame to be the light-crossing time of a causally connected region of size $\sim R/(\Gamma_j \Gamma_{co})$ (see, e.g., \citealp{narayanandkumar09}).
In the limit of $v_{co}=0$ ($\Gamma_{co}=1$), then $\Gamma_{em} = \Gamma_j$, $\theta = 0$, $\cos \alpha = \cos \Theta$, and we recover the ``usual" Doppler factor $\mathcal{D} = (\Gamma_j (1 - v_j \cos \Theta))^{-1}$, see eqs. (\ref{LF_minijets}), (\ref{lab_frame_angle}) and (\ref{Doppler_factor_minijets}).

As mentioned above, we will take $R \approx \Delta \Gamma_j^2$. The duration of the burst is $t_{GRB} \approx \Delta/c \approx R / (c \Gamma_j^2)$. For simplicity, we will assume that all minijets have the same Lorentz factor $\Gamma_{co}$.

\section{Anisotropic emission} \label{Section_3}

\subsection{Time delay} 

If minijets are isotropically distributed, minijets located along the line of sight ($\Theta=0$) are boosted towards the observer with a similar Doppler factor with ones at $\Theta = \Gamma_j^{-1}$; therefore, an observer sees emission within an angle $0<\Theta<1/\Gamma_j$.  However, for minijets emitted mainly at $\theta'=\pi/2$ (anisotropic minijets), their angle is $\theta = \Gamma_j^{-1}$ in the lab frame, so the Doppler boost of minijets at $\Theta = \Gamma_j^{-1}$ is larger by a factor of $\sim \Gamma_{co}^2 \gg 1$ than the one for minijets along the line of sight.  {\it Therefore, for anisotropic minijets, most of the emission arrives from a ``ring" of angle $\Theta = \Gamma_j^{-1}$, and thus, will be delayed compared with the isotropic emitter by}
\bea
t - t_0 \approx \frac{R - r \cos \Theta}{c} \approx \frac{R \Theta^2}{2 c} \approx \frac{\Delta}{2 c} \approx \frac{t_{GRB}}{2},
\eea
see eq. (\ref{time_delay}). This corresponds to the maximum time delay that can be experienced: lower level of anisotropy (smaller values of $n$) will correspond to shorter time delays, with the isotropic case showing no delay. 

\subsection{Steep decay}

The observed light curve -- from a conical source that is emitting isotropically and suddenly turns off -- declines quickly with time (e.g., \citealp{fenimoreetal95, kumarandpanaitescu00}), the so-called ``high latitude emission".  In this subsection, we will show that an anisotropic source decays even faster.  Instead of focusing on particular minijets, we will focus on the `envelope' emission, that is, the average emission of all minijets. We follow \cite{beloborodovetal11}, where they find that the luminosity is
\bea \label{luminosity_steep}
L \propto P(\theta') \Big(1 + \frac{t-t_0}{t_{GRB}}\Big)^{-(2+\beta)},
\eea
where $P(\theta')$ is the intrinsic (comoving) angular distribution of emission (per unit solid angle), normalized by $\int P(\theta') \, d\Omega = 4\pi$, and $\beta$ is the energy spectral index. Isotropic emission corresponds to $P(\theta')=1$ and yields a decay as $L \propto t^{-(2+\beta)}$, consistent with \cite{kumarandpanaitescu00}. 
For $\Gamma_{co} \gg 1$ ($v_{co} \approx 1$), the emission from a single minijet is narrowly beamed, therefore, the main factor that contributes to the anisotropy comes from their direction distribution, $P(\theta') \propto (\sin \theta')^n$. In this case, eq. (\ref{lab_frame_angle}) becomes 
\bea \label{angle_approx}
\cos \Theta \approx \frac{\cos \theta' + v_j}{1 + v_j \cos \theta'}.
\eea
For $r \approx R$ in eq. (\ref{time_delay}), we solve for $\cos \Theta$ as a function of time, and using eqs. (\ref{luminosity_steep}) and (\ref{angle_approx}), we find that the temporal decay index of the steep decay is\footnote{We note that eq. (\ref{angle_approx}) allows us to solve for $\cos \theta' = f$, therefore, $L \propto (\sin \theta')^n \propto [\sin (\arccos f)]^n = (1 - f^2)^{n/2}$, where $f$ is a function of time.} 
\bea
\delta = \frac{d \ln L}{d \ln t} = \Big(\frac{n}{2}\Big) \Bigg[\frac{\frac{t}{t-t_0} - \frac{t}{t_{GRB}}}{1+\frac{t-t_0}{t_{GRB}}} \Bigg] - (2+\beta) \Bigg[\frac{\frac{t}{t_{GRB}}}{1+\frac{t-t_0}{t_{GRB}}}\Bigg],
\eea
which at late times $t \gg t_0$ and $t \gg t_{GRB}$ approaches to
\bea
\delta = -(n/2) - (2 + \beta).
\eea
For a given spectral index, the more anisotropic the minijets are distributed, the steeper the emission will be. It also returns to well-known isotropic case for $n=0$. 

\section{Results} \label{Results}

In this section we present the results of our Monte Carlo simulations using the model described in Section \ref{Section_2}.  In the following, we use: $\Gamma_j = 100$, $\Gamma_{co} = 10$, $\theta_j = 0.05$, $\Delta = 3 \times 10^{11}$ cm and $R \approx 3 \times 10^{15}$ cm.  We simulate a large number of minijets ($\sim 10^6$). For simplicity, we assume that minijets produce Gaussian pulses in the observed light curve that appear at time $t-t_0$, with width $\delta t$, and we also assume that all minijets have the same luminosity in the comoving frame of the jet, $L_{mj}$, so that the peak of each pulse is $\mathcal{D}^4$ (we take $\beta=1$). We set the zero time to be $t_0 = 0$. 

\begin{figure}
\includegraphics[width=8.5cm, angle=0]{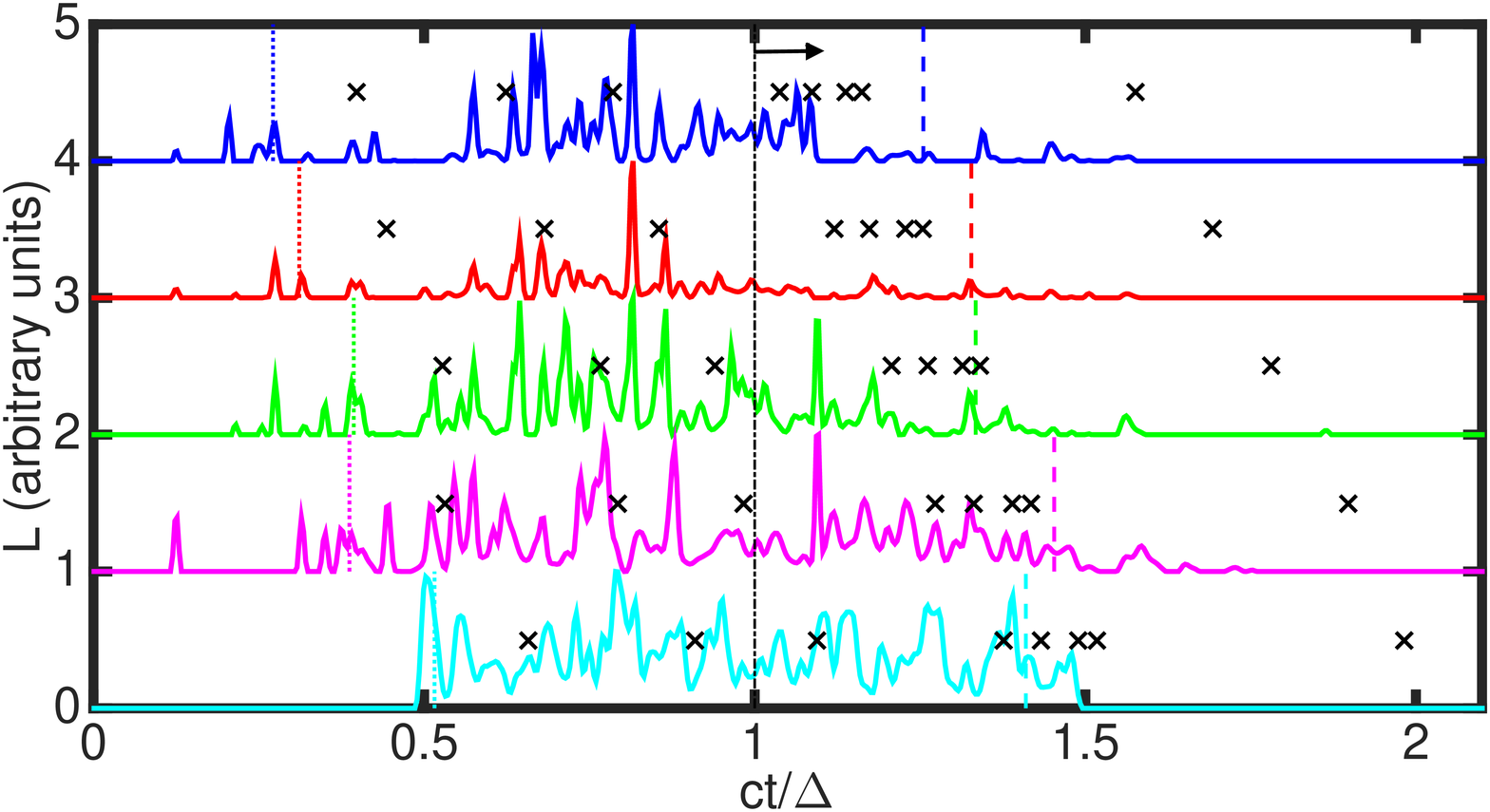}
\includegraphics[width=8.5cm, angle=0]{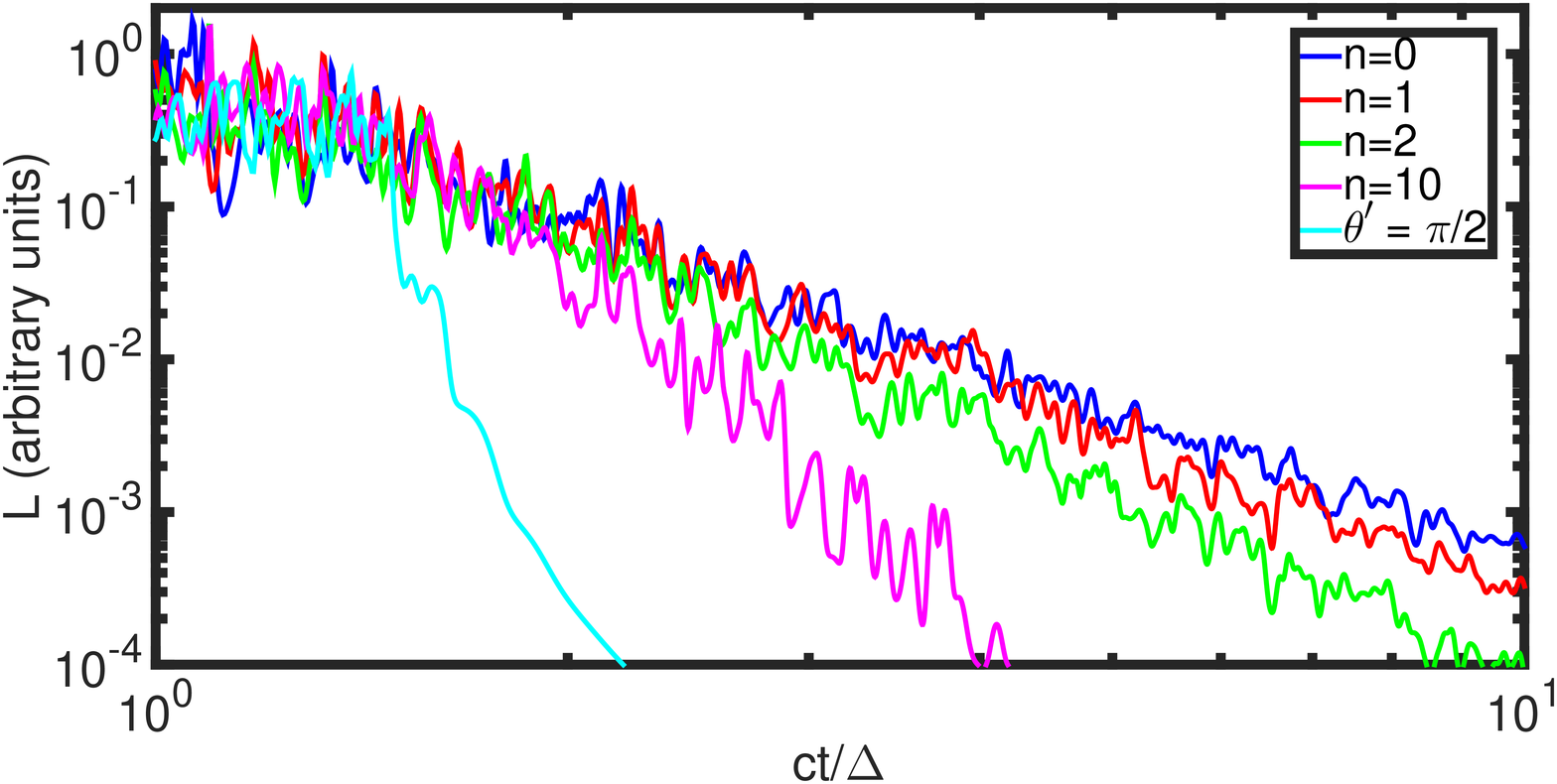}
\caption{Typical light curve for the minijet model; see Section \ref{Results} for a description of the used parameters. In both panels, the degree of anisotropy increases from top to bottom (see legend in bottom panel). {\it Top panel:} Linear scale plots, which show that anisotropy shifts the overall light curves to later times. Different light curves (normalized) have been shifted vertically for displaying purposes. The vertical dotted (dashed) lines for each light curve correspond to their values of $T_{05}$ ($T_{95}$), which are the times when 5 (95) per cent of all counts are accumulated by the detector (typically, the total duration of the emission is taken as $T_{90} = T_{95} - T_{05}$). $T_{05}$ generally increases as minijets are more anisotropic. A small background ``noise" has been subtracted to qualitatively mimic the real background subtraction in the detector.  We also include the observed peaks of the GeV light curves (black crosses) for the sample in Table \ref{table1}, scaled for each of the simulated light curves (see Section \ref{Section_GeV}). As the level of anisotropy increases, the peaks of the GeV light curves also shift to later times, making most of them (except for the case of GRB 080916C) consistent with $t \gae \Delta/c$ (marked with a dash-dotted line and an arrow). For one burst the peak time is outside the limits of the figure.
{\it Bottom panel:} Logarithmic scale, which shows that more anisotropic minijets yield faster decaying light curves during the ``high latitude" emission. Light curves have been arbitrarily shifted vertically to have the same approximate luminosity at $t \sim \Delta/c$.}    
\label{fig1}
\end{figure}

The properties of the anisotropic emission described in Section \ref{Section_3} can be seen in Fig. \ref{fig1}. We show the light curves (in a linear, and logarithmic scale) for different levels of anisotropy.  The isotropic case ($n=0$) and the case where all minijets point at $\theta' = \pi/2$ ($n \rightarrow \infty$) in the comoving frame are also shown.  The linear scale plot shows that, as the minijets' direction becomes more anisotropic (as $n$ increases), the light curve begins at a later time (see dotted lines close to $t_0=0$ for each case, which mark the approximate time when 5 per cent of the count are accumulated, $T_{05}$), while the logarithmic scale plot shows how the light curves become steeper at late times.  As can been seen, the $\theta'=\pi/2$ case yields a time delay of $\sim \Delta / (2 c)$ as predicted in Section \ref{Section_3}. 

Observationally, the fact that the time delay increases as anisotropy increases in Fig. \ref{fig1} is meaningless, since detectors will trigger only when enough counts are present. However, this has profound implications on the external shock light curves, which peak at $t \gae \Delta /c$, irrespective of the degree of anisotropy of the prompt emission. We will discuss this in Section \ref{Section_GeV}.

As explained above, an immediate consequence of a higher degree of anisotropy in the minijets is the fact that the late light curve, following the main emission episode shown in the top panel of Fig. \ref{fig1}, would decay more steeply than the isotropic case, and continues to show some level of variability \citep{narayanandkumar09}.  Here, we have quantified this decay in Section \ref{Section_3}, and our Monte Carlo simulations agree with our theoretical arguments.  The temporal decay for $n=0,1,2,10$ is expected to asymptotically be $\delta = -3, -3.5, -4, -8$, consistent with the bottom panel of Fig. \ref{fig1}.  The $\theta'=\pi/2$ case, which corresponds to an infinite value of $n$, simply drops to zero immediately after the prompt emission ends.

\section{Application} \label{Application}

\subsection{Time delay: GRB GeV peaks} \label{Section_GeV}

In the context of GRBs, GeV (100 MeV - 10 GeV) LAT (Large Area Telescope) light curves detected by the {\it Fermi} satellite last much longer than the variable prompt emission phase (50 - 300 keV) detected by GBM (Gamma-ray Burst Monitor), and decay as a power-law (e.g., \citealp{ackermannetal13}). These, and other pieces of evidence, point to their origin in the external forward shock (e.g., \citealp{kumarandbarniolduran09, zouetal09, ghisellinietal10}), which is produced as the relativistic GRB jet interacts with the circumburst medium (e.g., \citealp{sarietal98, wijersandgalama99, panaitescuandkumar00}). If the GeV emission during the prompt phase is also associated with an external forward shock origin, then it is natural to associate the peak of the GeV light curve, $T^{LAT}_p$, with the deceleration time of the blast wave.  This is the time when the external reverse shock has crossed the GRB ejecta and a rarefaction wave signals the external forward shock to steepen its deceleration profile, and thus, produces a break in the light curve (e.g., \citealp{sariandpiran95}).  In the simplest model, the deceleration time will occur at $\sim \Delta/c$, which corresponds to the end of the prompt emission, $\sim T^{GBM}_{90}$ (time for which 90 per cent, from 5 to 95 per cent, of the counts are detected), or after it, depending on if the reverse shock is relativistic or not (see, e.g., \citealp{sariandpiran95, sari97, panaitescuandkumar04} for a hydrodynamical treatment of the problem, and \citealp{mimicaetal09} for a full magnetohydrodynamical one). However, most of the observed GeV light curves rise and peak during the prompt emission phase, that is, the GeV light curves start decaying before the end of the prompt phase (e.g., \citealp{ackermannetal13}); see Table \ref{table1}.  This has been used to argue against their external forward shock origin (e.g., \citealp{heetal11, maxhametal11, beloborodovetal14}).  Here we argue that it is possible to circumvent this issue by assuming that the prompt emission is anisotropic.  In particular, we show that in the anisotropic minijets model, the prompt emission phase can be delayed by up to $\sim t_{GRB}/2$.  Therefore, the prompt emission is overall shifted in time (delayed) with respect to the peak of the GeV light curve, whose peak is always at $\gae \Delta/c$, which may result in a GeV peak {\it during} the prompt emission phase.

Given that each different simulated light curve has a different ``trigger" time, which is close to $\sim T_{05}$, we use the data in Table \ref{table1} to determine the corresponding peak time of the GeV light curve for each of the light curves (depending on their respective $T_{05}$ and $T_{90} = T_{95} - T_{05}$) approximately as $T_{peak} \approx T_{05} + (T^{LAT}_p/T^{GBM}_{90}) T_{90}$. We plot this $T_{peak}$ (normalized to $\Delta/c$) as black crosses in Figure \ref{fig1} for each of the light curves.  The GeV peaks shift to later times, to times $\sim \Delta/c$ or later, alleviating the concerns presented above.  The most extreme case is GRB 080916C, where even the extreme anisotropic case ($\theta' = \pi/2$) still yields a GeV peak long before $\sim \Delta/c$. For this case, it is possible that a smaller number of minijets is present and statistical fluctuations may play a significant role; see the end of Section \ref{Discussion}. 

\begin{table}
\begin{center}
\begin{tabular}{c|ccc}
\hline
GRB & $T^{LAT}_p$ [s] & $T^{GBM}_{90}$ [s] & $T^{LAT}_p/T^{GBM}_{90}$\\
\hline
080916C 	&	$	6.6 \pm 0.9 	$	&	$	64.22	\pm	0.8	$	&	$	0.1	$	\\
090323	&	$	40 \pm 30 	$	&	$	143.81	\pm	1	$	&	$	0.3	$	\\
090328	&	$	40 \pm 30 	$	&	$	65.65	\pm	2	$	&	$	0.6	$	\\
090510	&	$	0.9 \pm 0.1 	$	&	$	0.42	\pm	0.1	$	&	$	2.1	$	\\
090902B 	&	$	9 \pm 1 	$	&	$	22.18	\pm	0.3	$	&	$	0.4	$	\\
090926A 	&	$	11 \pm 2 	$	&	$	15.92	\pm	0.3	$	&	$	0.7	$	\\
091003	&	$	22 \pm 9 	$	&	$	21.07	\pm	0.4	$	&	$	1.0	$	\\
100414A 	&	$	20 \pm 10 	$	&	$	28.14	\pm	2	$	&	$	0.7	$	\\
110731A	&	$	4.9 \pm 0.7	$	&	$	7.54	\pm	0.3	$	&	$	0.6	$	\\
\hline
\end{tabular}
\end{center}
\caption{GRB, the peak of the LAT light curve ($T^{LAT}_p$) in the 100 MeV - 10 GeV band, the duration of the GBM prompt emission (50 - 300 keV band) light curve ($T^{GBM}_{90}$), and the ratio between these two values.  Most LAT light curves peak before the end of the prompt emission. (Data from \citealp{ackermannetal13}). } 
\label{table1}
\end{table}

\subsection{Steep decay: GRB early X-ray steep decay}

After their prompt emission phase, most GRBs show a period of rapid X-ray flux decay, which has been detected by XRT (X-ray Telescope) onboard the {\it Swift} satellite (e.g., \citealp{barthelmyetal05}).  This emission has been associated with either the high latitude emission (e.g., \citealp{kumarandpanaitescu00}) or the emission from a rapidly decaying central engine \citep{fanandwei05, barniolduranandkumar09}. As shown above, the high latitude emission from an anisotropic source decays faster than the isotropic one.  

When fitting the observed XRT light curves with a decaying power-law, a choice of the zero time must be made.  Usually, the zero time is selected to be the ``trigger" time of the BAT (Burst Alert Telescope) onboard {\it Swift},  and the decay index is $\alpha_{obs}(t_0=t_{trigger})$.  Alternatively, a zero time closer to the end of the prompt emission (closer to $T_{90}$) can also be chosen, as it would mark the real zero time of the ejection of the last ``shell" that produced the last spike in the prompt light curve, e.g., in the internal shock model.  The latter zero time yields less steep decay indices than the former.     

However, the {\it real} zero time of the emission for an anisotropic source should be shifted to a time that is {\it earlier} than the ``trigger" time of the detector. The trigger time in Fig. 1 is $\sim T_{05}$. As anisotropy increases, the zero time should be set to a time which is $\lae t_{GRB}/2$ earlier than the trigger time.  With this corrected zero time, $\alpha_{obs}(t_0=t_{trigger})$ will be steeper by only $\lae 20$ per cent (if the flux drops from $T_{90}$ to $\sim 10 T_{90}$), and this decay index can be compared with the theoretically expected one (see Section \ref{Section_3}). Many X-ray steep decay light curves decay faster than predicted by the isotropic case, which suggest that an anisotropic model is necessary to explain these cases (e.g, \citealp{cusumanoetal06, hilletal06, vaughanetal06})\footnote{Using the sample of 16 GRBs in \cite{obrienetal06}, which have an X-ray steep decay phase [$\alpha_{obs}(t_0=t_{trigger}) \gae 2.5$], 10 have similar $\gamma$-ray and X-ray spectra ($\beta_{\gamma} \sim \beta_X$), making them possibly consistent with the ``high latitude" emission interpretation. Out of these 10, 5 have slopes steeper than those predicted by the simple high latitude emission, that is, a large fraction of X-ray light curves decay faster than expected in this simple model.}.

\section{Discussion and Conclusions} \label{Discussion}

In the model presented in this Letter, the short timescale variability in the observed light curves is driven by compact emitting regions (``minijets") that move relativistically in the jet frame.  The presence of such emitters can result in jet variability shorter than the dynamical time at the central engine, as required to explain ultrafast flaring from blazars (e.g., \citealp{gianniosetal09, narayanandpiran12, giannios13}), and to explain fast variability in GRBs (e.g., \citealp{lyutikovandblandford03, kumarandnarayan09, lazaretal09}).

Physically, minijets can naturally develop if magnetic reconnection takes place in a Poynting-flux dominated jet. For jet magnetization $\sigma = B^2/(4 \pi \rho c^2) >1$, the reconnection outflows are expected to move with Lorentz factor $\Gamma_{co} \sim \sigma^{1/2}$ in the jet rest frame and in the plane of the current sheet (\citealp{lyubarsky05, giannios13}) and the orientation of the reconnection layers determines the directionality of the minijets.  Although the details of the orientation of the current sheets are currently uncertain and model-dependent, in general one expects the magnetic field lines in the jet to be toroidally dominated.  The end stage of magnetohydrodynamical instabilities responsible for dissipation in the jet may favor layers perpendicular to the jet propagation.  In addition, if the magnetic field changes polarity regularly at the central engine \citep{parfreyetal15}, then the reconnection layers will be naturally perpendicular to the jet propagation (in the jet rest frame).

Using a simplified model, we have calculated the observed light curves from minijets, randomly selecting their position within the jet. We consider an isotropic distribution of minijet directions and also consider more anisotropic distributions, for which the direction is more likely to be perpendicular to the direction of the jet (in the comoving frame of the jet).  We find that more anisotropic distributions yield: 1. observed light curves that start later, and 2. a high latitude emission that decays faster (see Fig. \ref{fig1}). 

Observationally, a light curve that starts later might seem inconsequential, since detectors ``trigger" whenever enough counts are detected.  However, this has deep implications for the peak of the external forward shock.  This peak is due to the deceleration of the GRB blast wave, which will decelerate as usual, whereas the prompt light curve will be shifted to later times (for anisotropic minijets), allowing for the deceleration to occur {\it during} the prompt emission phase. In the context of an external forward shock origin of the GeV light curves detected by {\it Fermi}, an anisotropic minijets model naturally explains why the GeV peak occurs during the prompt emission phase.  

If the GeV peak time, $T_p^{LAT}$ is indeed the deceleration time of the blast wave, then the jet bulk Lorentz factor $\Gamma_0$ can be estimated.  An anisotropic light curve can be delayed by, at most, $\sim t_{GRB}/2$, where $t_{GRB}$ is the duration of the GRB, usually measured as $T_{90}^{GBM}$.  In this case, the real deceleration time is $\lae T_p^{LAT} + T_{90}^{GBM}/2$, that is, a factor of $\lae 1 + 0.5 (T^{LAT}_p/T^{GBM}_{90})^{-1}$ larger.  However, since $\Gamma_0$ depends very weakly on the peak time, the estimates of $\Gamma_0$ (see, e.g., \citealp{ackermannetal13}) are not significantly modified.  

After the prompt phase, the high latitude emission takes over, and an anisotropic minijets model decays faster than the isotropic case.  The model can explain the GRB X-ray steep decay light curves that show a decay index steeper than that expected by the isotropic case.

We have used a Monte Carlo scheme to simulate the light curves in the minijets model.  The light curves do depend to some extent on the chosen parameters, in particular, on the number of minijets.  We have simulated a typical GRB light curve with $\sim 50$ pulses.  We verified that the main two features in this Letter remain unchanged as long as the number of simulated minijets is large.  However, low number of minijets produce light curves with small number of well-separated pulses, where the statistical uncertainty of their duration ($T_{90}$), their beginning ($T_{05}$) and their end ($T_{95}$) is much larger.  Comparing the peak time of the afterglow emission with $T_{90}$ in such bursts with few pulses may be misleading.

In this Letter, the dissipation radius has been chosen to be large, by requiring that the whole width of the jet comes in causal contact. Minijets could also dissipate at a smaller radius. In this case, the time delay expected from anisotropic minijets would be correspondingly smaller. However, both the GeV peaks (which occur during the prompt emission) and also the X-ray steep decay (steeper than the isotropic case) points to a large dissipation radius to explain the observations within the context of the anisotropic minijets model.

Our results point to a gamma-ray generation mechanism in GRBs that is anisotropic. Although the precise details of our model may vary, the overall features presented in this Letter are robust.  This provides an important step towards identifying the energy dissipation and radiation mechanisms taking place during the prompt emission phase, which continue to remain elusive.

\section*{Acknowledgements}

We thank Maria Petropoulou and Pawan Kumar for useful discussions. We acknowledge support from NASA grant NNX13AP13G.



\end{document}